\documentstyle[pra,twocolumn,aps]{revtex}

    \input epsf
    \begin{document}
    \draft
    \title{Dynamics and phase evolution of Bose-Einstein
condensates in one-dimensional optical lattices}
    \author{O.~Morsch$^{*}$, M.~Cristiani, J.H.~M\"uller,
D.~Ciampini,
    and E.~Arimondo}
    \address{INFM, Dipartimento di Fisica E. Fermi,
Universit\`{a} di Pisa, Via
    Buonarroti 2, I-56127 Pisa, Italy \\ $^{*}$ e-mail:
morsch@df.unipi.it}
    \date{\today}
    \maketitle
    \begin{abstract}
    We report experimental results on the dynamics and phase
    evolution
    of Bose-Einstein condensates in 1D optical lattices. The
    dynamical behaviour is studied by adiabatically loading
the
    condensate into the lattice and subsequently switching
off the
    magnetic trap. In this case, the condensate is free to
expand
    inside the periodic structure of the optical lattice.
The
    phase evolution of the condensate, on the other hand,
can be
    studied by non-adiabatically switching on the periodic
    potential. We observe decays and revivals of the
    interference pattern after a time-of-flight.
    \end{abstract}

    \narrowtext
    \section{Introduction}
    The study of Bose-Einstein condensates (BECs) in
periodic potentials has recently
    seen some major advances both on the
theoretical~\cite{javanainen99,trombettoni01,pedri01,kramer02}
and, most
    notably, the experimental side~\cite{orzel01,greiner02}.
Given the rapid
    progress made in optical lattices research with
ultra-cold
    atoms in the 1990s, it was only a matter of time before
the
    BEC community would extend those studies to the domain
of
    Bose-condensed atoms.

    So far, experiments with Bose-Einstein condensates have
    focused mainly on two areas. On the one hand, there have
been
    `conventional' investigations of the dynamics of
condensates in stationary,
    accelerated and pulsed optical
lattices~\cite{morsch01,hensinger01}. On the
    other hand, the very efficient loading of ultra-cold
atoms
    into a periodic structure afforded by BECs has also made
    possible the observation of quantum effects such as
number
    squeezing~\cite{orzel01} and the Mott-insulator
    transition~\cite{greiner02}.

    In previous experiments, we have already addressed the
issue of
    a BEC in an accelerated optical lattice and associated
    phenomena such as Bloch oscillations and Landau-Zener
    tunneling~\cite{morsch01,cristiani02}. In the present
work, however, we shall
    concentrate on the properties of BECs in stationary
periodic
    potentials. We have investigated both the dynamics and
the phase evolution of BECs in
    such potentials.

    This paper is arranged as follows. After briefly
presenting our
    experimental setup in section~\ref{setup}, we discuss
the
    results of our experiments on the free expansion of the
BEC in
    the lattice (section~\ref{free}). The phase evolution of
the
    condensate is the subject of section~\ref{phase}, which
is
    followed by conclusions and an outlook on further
experiments
    (section~\ref{conclusion}).
    \section{Experimental setup}\label{setup}
    Our experimental setup is described in detail
   elsewhere~\cite{morsch01,cristiani02}. Briefly, after
creating BECs of $N_0=1-2\times 10^4$
   $^{87}$Rb atoms in a triaxial time-orbiting potential
trap, we
   adiabatically lower the trap frequency
$\overline{\nu}_{trap}$ to the desired value and
   then superimpose onto the magnetic trap an optical
lattice
   (either along the vertical or the horizontal trap axis)
   created by two linearly polarized Gaussian laser beams
   intersecting at a half-angle $\theta$ and detuned by
$\approx 30\,\mathrm{GHz}$ above the
   rubidium resonance line. The periodic
   potential $U(z)=U_0\sin^2(\pi z/d)$ thus
created~\cite{footnote_directions} has a lattice spacing
   $d=\pi/[k \sin(\theta/2)]$, and the depth $U_0$ of the
potential
   (measured in lattice recoil energies $E_{rec}=\hbar^2
   \pi^2/2md^2$) can be varied between $0$ and $\approx
   20\,E_{rec}$ by adjusting the laser intensity using
   acousto-optic modulators. In our setup, we could realize
a horizontal lattice with $d=1.56\,\mathrm{\mu m}$
   and a vertical lattice with $d=1.2\,\mathrm{\mu m}$. The
number of lattice sites occupied
   by the condensate lay between $10$ and $15$, depending on
the
   trap frequency and the lattice spacing. When using the
vertical
   lattice with the magnetic trap switched off (in the
expansion
   experiments, see below), we accelerated the lattice
downwards
   by chirping the frequency difference between the lattice
beams
   in order to offset gravity in the rest-frame of the
lattice.

   \section{Dynamics: Free expansion inside a 1D
   lattice}\label{free}
\begin{figure}
    \centering\begin{center}\mbox{\epsfxsize 2.8 in
\epsfbox{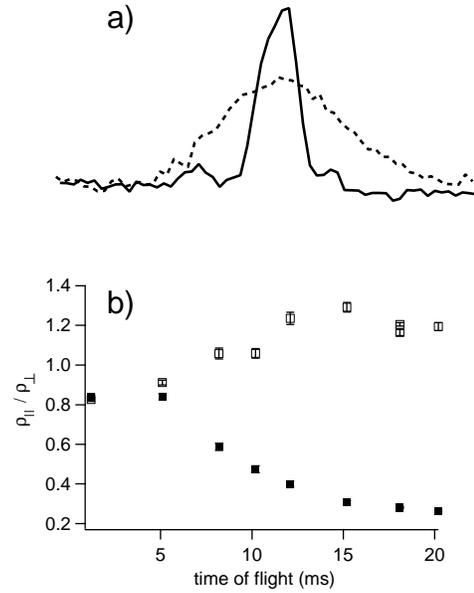}}
    \caption{Effect of the one-dimensional optical lattice
on the condensate expansion. In (a), the
    difference in condensate width along the lattice
direction after $\approx 23\,\mathrm{ms}$ of free expansion with
(solid line)
    and without the lattice (dashed line) is clearly
visible. The evolution of the aspect ratio of the condensate
    in a lattice with $U_0=20\,E_{rec}$ is shown in (b).
When the lattice is present (filled symbols), the condensate does
not expand in the
    lattice direction and hence the aspect ratio decreases.
Without the lattice (open symbols), for the same magnetic trap the
    aspect ratio increases.}\label{Fig_asp_lasphys}
    \end{center}\end{figure}

   We studied the dynamics of the condensate in a stationary
   lattice by observing the free expansion of a condensate
   adiabatically loaded into the lattice. In order to load
the condensate into the optical lattice, after
   adiabatically reducing the mean magnetic trap frequency
${\overline{\nu}}_{trap}$ to the
   desired value we linearly increased the lattice depth
from $0$
   to $U_0$ in a time $t_{ramp}=150\,\mathrm{ms}$. Since
typically
   the chemical potentials $\mu_0$ of our condensates in
traps with
   frequencies ${\overline\nu}_{trap}/2\pi\approx
20-70\,\mathrm{Hz}$ lie between $50\,\mathrm{Hz}$
   and $200\,\mathrm{Hz}$, the adiabaticity
condition~\cite{band02}
   $t_{ramp}>h/\mu_0$ was satisfied even for small
   chemical potentials. After that, {\em only} the magnetic
trap was switched
   off and the condensate was imaged after a variable time
   $t_{exp}$ of free expansion inside the optical lattice.
The effect of a deep lattice on the
   free expansion is clearly evident in  the condensate
images of Fig.~\ref{Fig_asp_lasphys} (a) and in the measurements
of the condensate aspect ratio $\rho_{||}/\rho_{\perp}$ shown in
Fig.~\ref{Fig_asp_lasphys} (b) ($\rho_{||,\perp}$ denoting the
$e^{-1}$ half-widths of Gaussian fits to the density profile in
the lattice direction and perpendicular to it, respectively). As
in the presence of the lattice the condensate does not expand at
all in the lattice direction, the aspect ratio drops off sharply
with increasing time-of-flight as the BEC expands in the (unbound)
perpendicular direction. The lack of expansion in the lattice
direction reflects the fact that the condensate
   has effectively been split up into several smaller
condensates confined in the individual lattice wells. In the
perpendicular direction we have observed an enhanced expansion
when the lattice is present, which can be explained by the
increase in the chemical
   potential when the lattice is ramped up (as calculated by
Pedri {\em et al.}~\cite{pedri01}).

\begin{figure}
    \centering\begin{center}\mbox{\epsfxsize 2.8 in
\epsfbox{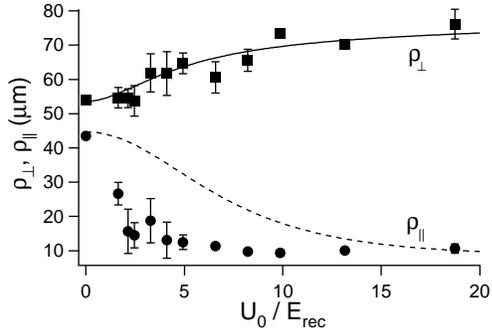}}
    \caption{Dependence on the lattice depth of the
condensate widths after $22\,\mathrm{ms}$ time-of-flight.
    The initial size (including the $5\,\mathrm{\mu m}$
resolution of our imaging system)
     of the condensate is $\approx 10\,\mathrm{\mu m}$. In
the lattice direction, for lattice depths $>2\,E_{rec}$
     the condensate does not expand anymore. This early
cut-off is not predicted by our simple
     scaling approach using the effective mass (dashed
line). The expansion in the perpendicular direction,
     however, is reproduced well by this model (solid line)
when the increase in chemical potential
     with $U_0$ is taken into account (see text). In this
experiment the mean trap frequency was $25\,\mathrm{Hz}$,
     with $\nu_{\perp}/\nu_{||}=\sqrt{2}/1$.}\label{Fig_depth_lasphys}
    \end{center}\end{figure}

   We modeled the expansion of the condensate in the
presence of the 1D lattice using the equations
   derived in~\cite{castin96,kagan96} with a slight
   modification along the lines of Ref.~\cite{kramer02}: In
the ansatz leading to the differential
   equation of~\cite{castin96}
for the scaling
   factor $\lambda_{||}(t=t_{exp})=
\rho_{||}(t=t_{exp})/\rho_{||}(t=0)$ along the lattice direction
   we replaced the atomic mass $m$
   by the effective mass $m^*(U_0)$, introduced
in~\cite{kramer02}
   in analogy with a solid state physics approach, as
derived from a band structure
   calculation for a periodic potential of depth $U_0$. As
can be seen in Fig.~\ref{Fig_depth_lasphys}, taking into account
the variation of the chemical potential with $U_0$ this model
reproduces well our experimental data for the perpendicular
expansion of the condensate (the theoretical plots are corrected
for the $5\,\mathrm{\mu m}$ resolution of our imaging system).

    For the lattice direction, this
   approach gives the correct result for deep lattices
   ($U_0\gg E_{rec}$), for which intuitively one expects the
condensate to be broken up and the
   individual parts to be essentially confined to the
lattice wells, suppressing the expansion
   along the lattice direction. The
   {\em qualitative} behaviour in the intermediate regime
   ($0<U_0<5\,E_{rec}$) is also reasonably well reproduced.
We find, however, that the experimental expansion along the
lattice direction is considerably less
   than the theoretical prediction in the intermediate
regime. We have checked that taking into account the finite
momentum spread of the condensate when
   calculating the effective mass only leads to a correction
on the percent level and thus cannot explain the deviation of our
   experimental data from the numerical calculations
neglecting
   mean field corrections. This might indicate that there
are effects
    such as self-trapping~\cite{trombettoni01} due
   to the mean-field interaction that further reduce the
expansion in the lattice direction. Another hint in this direction
is given by the fact that when we vary the number of atoms $N$ in
the condensate, for large $N$ the width in the lattice direction
of the expanded condensate actually starts to decrease rather than
increase. Again, this might be explained by mean-field effects
which become important in the dynamics of the condensate expansion
when $N$ gets large, with non-linear effects like self-trapping
reducing the observed width $\rho_{||}$.

As described in detail in~\cite{morsch02}, we could also directly
deduce from our data the dependence of the
   chemical potential on $U_0$ from the initial
perpendicular size
   $\rho_{\perp}(t=0)$ of the condensate in the presence of
the lattice inferred from
   the size $\rho_{\perp}(t=t_{exp})$ measured after an
expansion time $t_{exp}$ and the ratio
   $\lambda_{\perp}(t=t_{exp})/\lambda_{\perp}(t=0)$ of the
scale factor
   $\lambda_{\perp}(t)$. The chemical potential thus
measured
   agrees well with the prediction of~\cite{pedri01}.

\begin{figure}
    \centering\begin{center}\mbox{\epsfxsize 2.8 in
\epsfbox{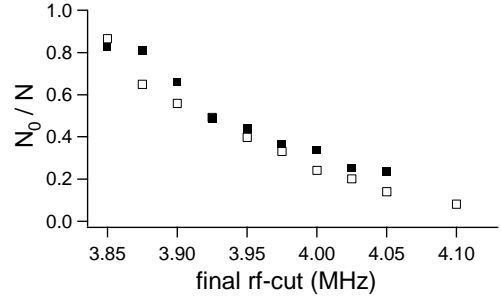}}
    \caption{The condensate fraction $N_0/N$ with (filled symbols)
    and without (open symbols) optical
    lattice as a function of the final rf-cut in the
evaporation. When the lattice is present, the condensate fraction
is
    comparable to or even slightly larger than without the
lattice. In this experiment, $U_0\approx 15\,E_{rec}$
    and the mean trap frequency was
$26\,\mathrm{Hz}$.}\label{Fig_condfrac_lasphys}
    \end{center}\end{figure}

   Finally, we note here that in the limit of large lattice
depths, the adiabatic loading
   of a BEC into and optical lattices effectively realizes
an adiabatic transformation
   between a 3D condensate and an array of 2D condensates.
The condition $\mu_{3D}<\hbar \omega_{lat}$ of
Ref.~\cite{gorlitz01}
    (where $\omega_{lat}= 2(E_{rec}/\hbar)\sqrt{U_0/E_{rec}}$
is the
   harmonic approximation for the oscillation frequency in a
lattice well) for
   the condensates in each well to be in the 2D limit is
always satisfied for the small number of atoms
   in a single well ($\approx 10^3$) present in our
experiment. For an
   array of 2D condensates obtained by creating the
condensate in the combined potential
    of the harmonic trap and the lattice, Burger {\em et
al.}~\cite{burger02} have shown that
   in the case of their cigar-shaped condensate (with the
long axis
   along the lattice direction), the transition temperature
$T_c^{2D}$ in the
   presence of the lattice is significantly lower than
$T_c^{3D}$ in the 3D
   case (i.e. in the magnetic trap without the lattice).
   Calculating the critical temperature $T_c^{2D}$ along the
same lines
   for our system, we find that $T_c^{2D}\approx T_c^{3D}$
due to the larger number of atoms per
   lattice site in our geometry, and
   hence we expect no significant change in the condensate
fraction in the
   presence of the lattice. In fact, experimentally we even
find a consistently
   larger condensate fraction after ramping up the lattice,
as seen in Fig.~\ref{Fig_condfrac_lasphys}. This
   result indicates that, with an appropriate choice of
parameters, a 1D optical lattice could be used to
   investigate adiabatic transformations between 3D and 2D
   condensates which could, e.g., be exploited to create
   condensates from thermal clouds by changing the
dimensionality
   of the system, as recently proposed by Olshani and
   Weiss~\cite{olshanii}, similarly to the change in the shape of the potentials in
Refs~\cite{stamperkurn99,pinkse97} for other
   geometries.

   \section{Phase evolution: Behaviour of the interference
   pattern}\label{phase}
\begin{figure}
    \centering\begin{center}\mbox{\epsfxsize 2.8 in
\epsfbox{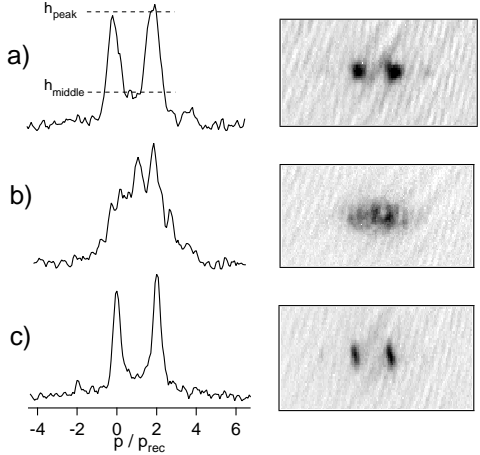}}
    \caption{Evolution of the interference pattern of a
condensate released from an optical lattice after non-adiabatic
    loading (absorption images and integrated profiles). The
distinct two-peaked structure visible immediately after loading
(a)
    washes out after a few milliseconds (b). For long
holding times in the lattice, the initial structure
    reappears (c). In this experiment, the mean trap
frequency was $26.7\,\mathrm{Hz}$ and the lattice
    depth $U_0\approx 15\,E_{rec}$. The holding times in the
lattice from (a) to (c) were $1$,$22$ and
    $300\,\mathrm{ms}$, respectively. In (a), the definition
of the visibility is illustrated.}\label{Fig_visi_lasphys}
    \end{center}\end{figure}

   In order to study the phase evolution of a condensate in
an
   optical lattice, we performed a {\em non-adiabatic}
loading
   procedure, meaning that the condensate initially was not
in the
   ground state of the combined potential of the magnetic
trap and
   the optical lattice. In a typical experiment, the optical
lattice was ramped up in
   $\tau_{ramp}\approx 1-5 \,\mathrm{ms}$, after which the
   potential was kept at its maximum value $U_0$ for a
holding time
   $t_{hold}$. At the end of the holding time, the lattice
was accelerated in $1\,\mathrm{ms}$ to one (lattice) recoil
velocity by
   chirping the frequency difference between the lattice
beams. Immediately after that,
   both the lattice and the magnetic trap were switched
   off.  After a time-of-flight of $20-22\,\mathrm{ms}$, the
   expanded condensate was imaged using a resonant probe
flash.
   Figure~\ref{Fig_visi_lasphys} shows typical absorption
images and integrated lines profiles obtained in
   this manner for different holding times. For short
   times, a clean double-peak structure is visible, as
expected
   from the interference between the condensates expanding
from
   the individual lattice wells (with a $\pi$ phase
difference
   between them due to the final acceleration). During the
first
   few milliseconds, this pattern evolves into a more
complicated
   structure featuring several additional peaks, and finally
   washes out completely, resulting in a single
   Gaussian-shaped lump. For long waiting times, the
two-peaked
   structure reappears.

   Intuitively, these results can interpreted as follows.
   Initially, the condensate is broken up into several
pieces when
   the optical lattice is abruptly switched on, and the
   condensate fragments are locally compressed. This leads
to a
   different (local) chemical potential and hence a
different
   phase evolution at the individual lattice sites. As a
result,
   the interference pattern after a time-of-flight evolves
from a
   clean two-peaked structure (all condensate fragments in
phase)
   into a broad Gaussian distribution, reflecting the fact
that the
   individual fragments have accumulated different phases.
For
   long holding times, a common phase is re-established
through
   tunneling between adjacent lattice sites and dissipation
of
   energy into higher-lying modes (more detailed studies of
these
   phenomena are under way).

\begin{figure}
    \centering\begin{center}\mbox{\epsfxsize 2.8 in
\epsfbox{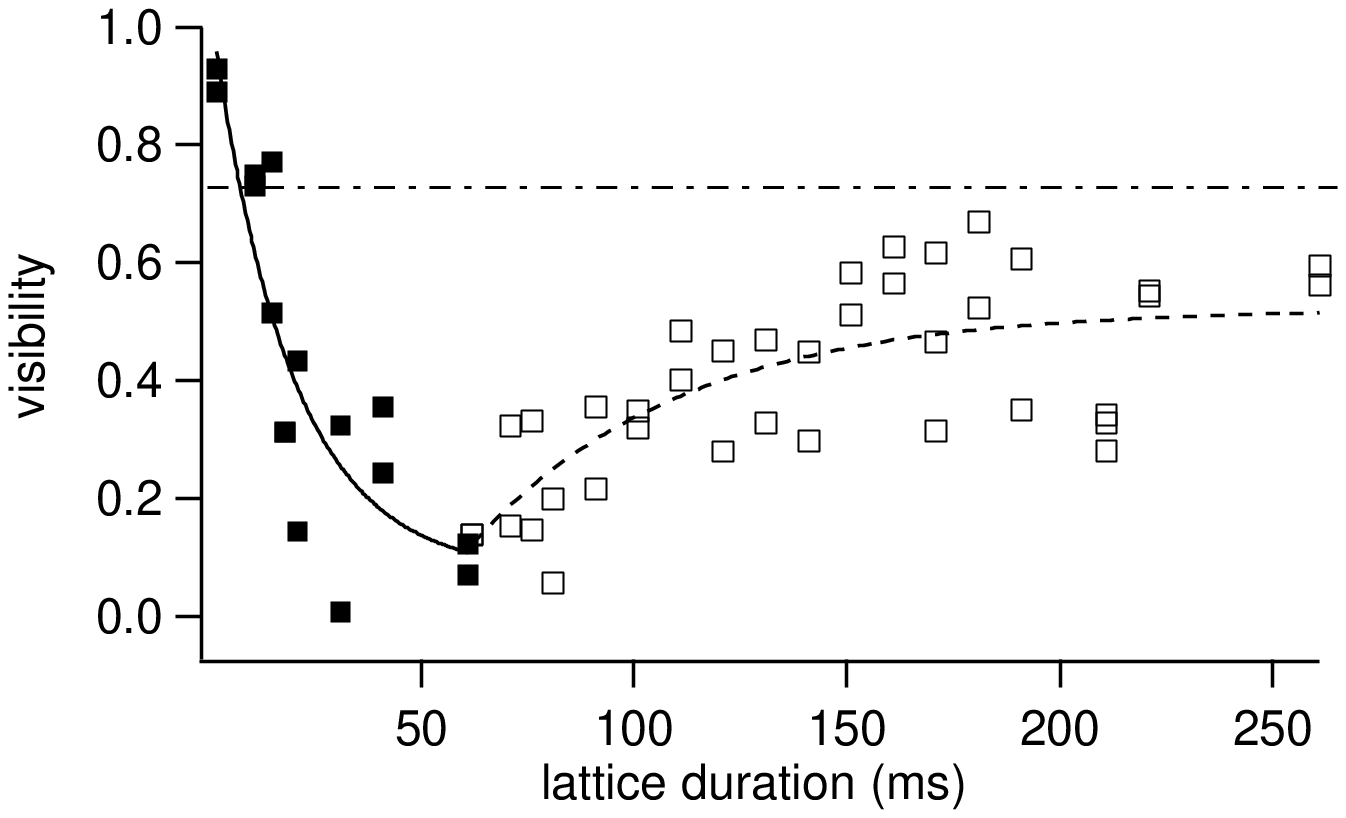}}
    \caption{Decay and revival of the interference pattern.
After the initial dephasing in a lattice
    with $U_0\approx 12\,E_{rec}$ (filled symbols), the
lattice depth was lowered to $\approx 4\,E_{rec}$ (open symbols),
resulting
    in a faster revival of the visibility. The solid and
dashed lines are exponential fits with time constants
    $\tau_{dephase}\approx 18\,\mathrm{ms}$ and
$\tau_{rephase}\approx 50\,\mathrm{ms}$, respectively. The
    dash-dotted line represents the visibility observed for
a condensate {\em adiabatically} loaded
    into the lattice with a ramp time
$\tau_{ramp}=150\,\mathrm{ms}$.}\label{Fig_visrev_lasphys}
    \end{center}\end{figure}

   In order to quantify the degree of coherence between the
lattice sites, we characterized the interference pattern
(integrated perpendicular to the lattice direction)
    through
   a visibility $\xi$ calculated as
   \begin{equation}
  \xi=\frac{h_{peak}-h_{middle}}{h_{peak}+h_{middle}},
   \end{equation}
   where $h_{peak}$ is the mean value of the absorption
image at
   the position of the two peaks, and $h_{middle}$ is the
value of
   the absorption image mid-way between the two peaks
(averaged
   over a range of $1/5$ of the peak separation) as
illustrated in Fig.~\ref{Fig_visi_lasphys} (a).
   In a typical experiment, $\xi$ initially rapidly
decreases
   from a value of $\approx 0.6-0.9$ to roughly $0$ within
   $\approx 5-20\,\mathrm{ms}$, depending on the trap
frequency
   $\overline{\nu}_{trap}$ (see
Fig.~\ref{Fig_visrev_lasphys}). If the lattice depth is kept
constant throughout, for trap frequencies
${\overline{\nu}}_{trap}>30\,\mathrm{Hz}$ the visibility then
typically rises
   again and begins to fluctuate in an apparently random
manner
   for a few tens of milliseconds. Finally, these
fluctuations die
   out and $\xi$ stabilizes at a value close to the initial
   visibility.

   For weak magnetic traps with
   ${\overline{\nu}}_{trap}<30\,\mathrm{Hz}$, after the
initial
   decay in visibility $\xi$ stays close to $0$ for up to
$200\,\mathrm{ms}$. Under these conditions, we could induce a
faster revival
   of the interference pattern by lowering the optical
lattice
   depth and hence increasing the tunneling rate between
adjacent
   lattice wells. Fig.~\ref{Fig_visrev_lasphys} shows the
initial
   decay for a lattice depth of $12\,E_{rec}$ and the
subsequent
   revival when the lattice is lowered to $\approx
4\,E_{rec}$.

   The interpretation of the phase evolution is complicated
by the
   fact that when the lattice is switched on
non-adiabatically,
   radial oscillations of the condensate are also excited.
These
   oscillations damp out after a few cycles. Also, the
temperature
   of the condensate initially rises and levels out at a
higher
   value. In our experiments, $T/T_c$ (where $T_c$ is the
critical
   temperature for condensation) typically rises from
$\approx
   0.7$ to $\approx 0.85$. This re-thermalization, which
takes
   places on a timescale that is comparable both to the
damping
   time of the radial oscillations and the revival time of
the
   visibility at fixed lattice depth, may take place through
a
   re-distribution of energy into higher-lying modes of the
   condensate. Theoretical studies of these processes are
   currently being carried out.
   \section{Conclusions and outlook}\label{conclusion}
   In a stationary optical lattice, both dynamical and phase
   properties of a BEC can be studied. By adiabatically
loading
   the BEC into the lattice, we have studied its free
expansion
   after switching off the magnetic trap. From the expansion
   perpendicular to the lattice direction we were able to
deduce
   the chemical potential of the BEC in the lattice as a
function
   of the lattice depth. The expansion in the lattice
direction
   for small lattice depths still needs further
investigation as
   it is not well reproduced by a simple treatment involving
the
   effective mass. Non-linear effects might come into play,
   possibly reducing the expansion through self-trapping.

   The phase evolution of a condensate non-adiabatically
loaded
   into a lattice is another intriguing area of study that
   deserves closer investigation. We have observed an
initial
   decay of the interference pattern followed, for long
holding
   times, by an almost complete revival. The role of
tunneling in
   this revival is evident from the fact that when the
lattice
   depth is lowered after the initial dephasing period, a
revival
   in the visibility is induced. The interplay between
tunneling,
   re-thermalization and damping of the radial oscillations
   excited by the non-adiabatic switch-on of the lattice
needs to
   be looked at more closely in future investigations. In
   particular, the role of the non-condensed fraction needs
to be
   elucidated.

   \section*{Acknowledgments}
   The authors wish to thank P.B.~Blakie, C.J.~Williams,
   P.S.~Julienne, S.~Stringari and Y.~Castin for useful
   discussions. This work was supported by the MURST
(PRIN2000
   Initiative), the INFM (Progetto di Ricerca Avanzata
   `Photonmatter'), and by the the EU through the Cold
Quantum
   Gases Network, Contract No. HPRN-CT-2000-00125. O.M.
gratefully
   acknowledges a Marie Curie Fellowship from the EU within
the IHP
   Programme.

    \newpage

    \newpage

    \newpage

    \newpage

     \newpage


\begin{references}
\bibitem{footnote_directions}For simplicity, in the
following we shall
    assume that the lattice direction is along $z$. The
different
    frequencies along the lattice direction and
perpendicular to
    it will be denoted by $\nu_{||}$ and $\nu_{\perp}$,
    respectively.
\end{references}
    \end{document}